# Fluence compensation for real-time spectroscopic photoacoustic imaging

MinWoo Kim, Geng-Shi Jeng, Matthew O'Donnell, Ivan Pelivanov

Department of Bioengineering, University of Washington, Seattle, WA, 98105, USA

**Abstract**. Recently we demonstrated an integrated photoacoustic (PA) and ultrasound (PAUS) system using a kHz-rate wavelength-tunable laser and a swept-beam delivery approach. It irradiates a medium using a narrow laser beam sweeping at high repetition rate over the desired imaging area, in contrast to the conventional PA approach using broad beam illumination at a low repetition. One significant advantage of this approach is that the fundamental problem of decoupling local light absorption at a point from optical fluence at the same point can be solved. Here, we present a fluence compensation method and demonstrate its performance in phantom studies. We adopted analytic fluence models, extending diffusion theory for the case of a pencil beam obliquely incident on a medium, and developed robust methods to estimate medium optical parameters using PA measurements acquired from multiple irradiation positions. We conducted comprehensive simulation tests and phantom studies using well-known contrast-agents to validate the reliability of the fluence models and spectral corrections.

## 1 Introduction

PA multispectral imaging relies on the known optical absorption spectrum of target chromophores. The PA signal, however, depends not only on the optical absorption spectrum of a target but also on the wavelength-dependent optical fluence at the same site. A critical challenge for in vivo, quantitative PA spectroscopic imaging is predicting light transport noninvasively in a turbid medium [1, 2]. The spatial distribution of fluence depends on the optical absorption and scattering properties of that medium, but they are usually unknown in advance. Also, since light illumination at a target position is nonuniform over wavelength, the medium induces spectral distortion between the nominal PA-measured optical absorption spectrum and the true target's spectrum. Therefore, for true quantitative spectroscopic (or multi-wavelength) PA imaging capable of identifying molecular constituents in the medium under study, the optical fluence must be simultaneously estimated and compensated. This requires not only an appropriate fluence model, but also an effective method to estimate the fluence distribution at each wavelength from PA measurements.

Many different fluence correction methods have been proposed for a homogenous scattering medium such that the light distribution can be closely represented by simple formulas or mathematical expressions based on prior knowledge of the medium's optical properties [1, 3-6]. Unfortunately, none translate into clinical tools. Indeed, optical constants reported in the literature may vary a few orders of magnitude depending on the measurement technique, tissue condition, and geometry [7]. Background tissue properties may also change dynamically based on tissue blood content and blood oxygenation level. Optical constants measured even with the same device, under the same experimental conditions, and for the same person may vary. Furthermore, these properties may change during medical procedures and interventions. Thus, tissue optical properties must be estimated during PA imaging for accurate spectroscopic measurements.

Some research groups adopted other spectroscopic modalities to estimate the optical parameters needed for analytic fluence models [8, 9]. Estimating fluence from PA measurements without the help of other tools and at near real-time rates, however, is highly desirable for clinical applications.

In Ref. [3], point source tissue illumination from different positions along the tissue surface was proposed to calculate the laser fluence distribution within the medium and then apply the estimated fluence for PA spectral decomposition. However, the authors did not show how to integrate this method into real-time PA scanners since broad beam illumination is most commonly used in PA imaging systems [3]. Leveraging the results of this study, we recently introduced a real-time interleaved photoacoustic-ultrasound (PAUS) fast-sweep scanner [10], where unlike previous delivery systems coupling laser pulses into all fibers in a bundle simultaneously, light is coupled into individual fibers **sequentially** (see Fig. 1), but at a very high rate.

A unique diode-pumped wavelength tunable (700 nm – 900 nm) laser emitting about 1 mJ pulses at 1000 Hz, with wavelength switching in less than 1 ms for any arbitrary wavelength order, was designed especially for the fast-sweep PAUS scanner. To maximize exposure, we illuminate with a narrow (~ 1 mm in diameter) beam and switch it from fiber-to-fiber at 1000 Hz, resulting in one loop around the US probe per single-wavelength frame in only 20 ms (Fig. 1). The next loop uses another wavelength without delay; the procedure repeats over all wavelengths. That is, instead of illuminating with a broad beam, we use fast-scanning (or fast-sweep) over the same illumination area.

In our PAUS system, ten fibers are uniformly spaced along each elevational edge of the US

array (e.g. 20 fibers in total, Figs. 1, 2). Every laser shot from a single fiber is followed by a sub-image reconstruction, i.e. 20 sub-images in all, which are then coherently summed to form the full PA frame. The kHz rate enables 50 full 'loops' of the laser beam around the probe per second, resulting in a 50 Hz PA frame rate. For stable spectral decomposition, 10 wavelengths (i.e., 700, 715-875 nm every 20 nm) form the spectroscopic sequence. More details on the system can be found in Ref [10].

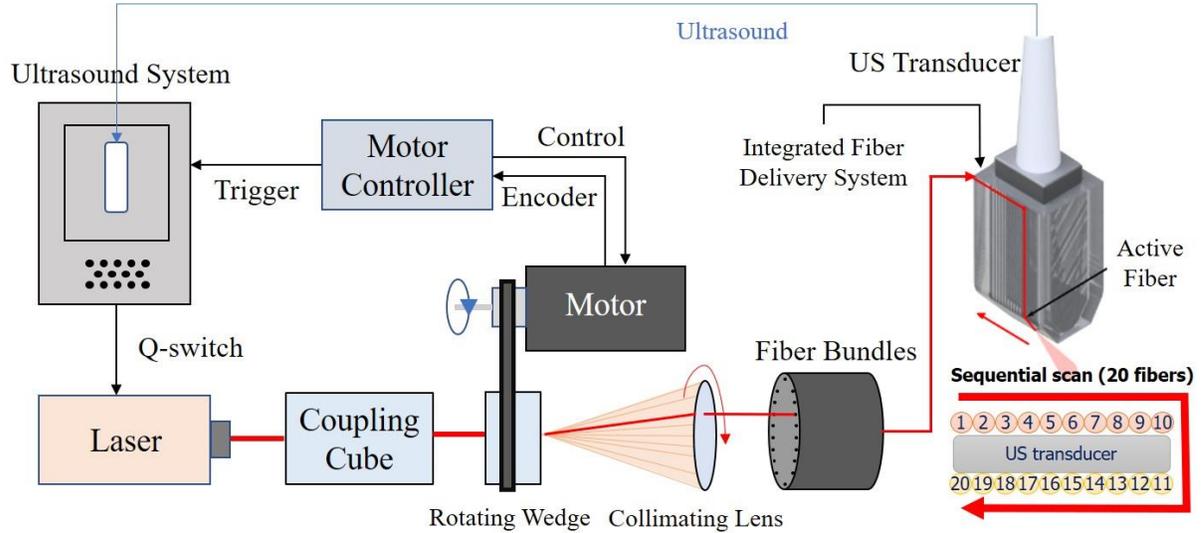

**Fig. 1.** Real-time integrated photoacoustic and ultrasound (PAUS) system used in these studies. The ultrasound system programmably controls the laser, motor controller, and US transducer. The motor controller synchronizes emission with the centers of 20 fibers in the bundle, delivering a trigger to the US system when properly aligned for each fiber. The US system then externally triggers the compact laser, transmitting a pulse at about a 1 kHz repetition rate with a wavelength switchable from pulse-to-pulse over the range of 700 nm to 900 nm. With absolute position control, a precise rate is not needed for external laser triggering, ensuring maximal light delivery to each fiber. Motor speed variations only slightly alter the overall frame rate of 50 Hz. A total of 20 fibers are arranged on two sides of the linear array US transducer, as shown in the zoomed front view in the bottom right corner. The system interleaves laser beam and focused US transmissions for interleaved PA spectroscopic imaging and US B-mode imaging.

Here we describe in detail how to use partial PA images from every fiber to estimate laser fluence. Indeed, when light emerges from different fibers, it propagates different distances to a target and, therefore, the amplitude of a partial PA image reconstructed from a single fiber illumination will depend on that fiber's position around the US probe (see Fig. 2). For every fiber light source, we adopted analytical fluence models extending diffusion theory for narrow beam (pencil beam) illumination obliquely incident on a semi-infinite homogeneous scattering medium [11]. Using this model, we explore methods to extract optical parameters to assess the fluence distribution within the medium using data acquired at different fiber positions and wavelengths. In particular, estimated parameters include the effective light attenuation $\mu_{eff}$, and reduced light scattering $\mu_s'$ coefficients of a turbid medium.

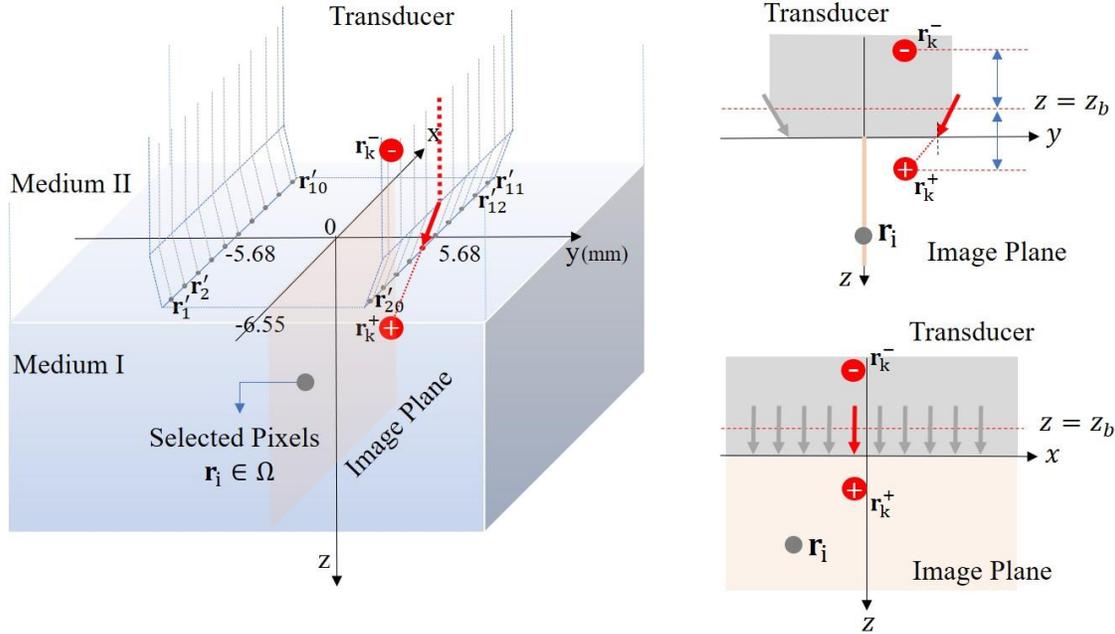

**Fig. 2.** Schematic diagram of laser sources and bulk media. Each fiber is mounted on the side of a linear-array transducer in an ambient medium (Medium II), and irradiation at the tip of the fiber into a scattering medium (Medium I) is modeled as a pencil beam. The center position of the face between the transducer and Medium I is (0,0,0). The image plane is (y=0). The y-directional position of every irradiation point is either 5.68 mm or -5.68 mm, and the tilt angle of the laser beam is $\theta = 35°$. The pencil beam can be represented as two isotropic point sources satisfying the boundary condition that light propagation from $z < z_b$ into $z > z_b$ is approximately 0. The extrapolated boundary $z = z_b$ is determined by the refractive indexes of the two media and the transducer. The sign of imaginary sources indicates their polarity. Source positions are determined by the transport mean free path $l_t$.

Building on the basic analytic model, we then explore robust methods to estimate these parameters from noisy data. Simulations were conducted to clarify quantitative errors arising from adopted optical fluence models, variations in medium properties, and measurement noise levels. Finally, we demonstrate the reliability of the correction methods via phantom studies.

## 2   Model

*2.1 Photoacoustic Signal*

Acoustic pressure obtained for the $k$th optical fiber, optical wavelength $\lambda_j$ at discrete position $\mathbf{r_i}$ can be denoted as

$$p_{j,k}(\mathbf{r}_i) = \Gamma \bar{\mu}_a^{(j)}(\mathbf{r}_i)\Phi_{j,k}(\mathbf{r}_i), \tag{1}$$

where $k \in \{0,1,\ldots,19\}$ and $j \in \{0,1,\ldots,9\}$, $\Gamma$ is the Grünesien parameter, $\bar{\mu}_a$ is the optical absorption coefficient and $\Phi$ is the light fluence. We assume here that the parameter $\Gamma(\mathbf{r})$ governing PA efficiency is constant over space $\{\mathbf{r}|\mathbf{r} = (x,y,z), \mathbf{r} \in \Omega\}$ and wavelength. This assumption is not critical for relative spectroscopic measurements, but is important for absolute concentration estimates of specific chromophores producing the PA signal.

The absorption spectrum is given as

$$\bar{\mu}_a^{(j)}(\mathbf{r}_i) = \sum_{l=1}^{L} \alpha_l^{(j)} C_l(\mathbf{r}_i), \tag{2}$$

where $L$ is the number of chromophore (absorber) types and $C_l$ and $\alpha_l^{(j)}$ are the concentration and absorption coefficient of the $l$th chromophore type, respectively. The ultimate goal of quantitative spectroscopic PA imaging is to estimate the relative, or absolute, concentration of a particular chromophore of interest at each position within the imaging field from measurements given the known absorption spectrum $\alpha_l^{(j)}$ of that chromophore. Thus, correct modeling and computation of the wavelength-dependent optical fluence distribution $\Phi_{j,k}(\mathbf{r}_i)$ at every source position is required to guarantee accurate quantification of the target chromophore.

*2.2 Optical Fluence Model*

Photon transport in tissue can be modeled with the radiative transfer equation (RTE), also called Boltzmann's transport equation. Since PA imaging is associated with a pulsed light source, the time-independent form is applicable to time-integrated quantities as

$$q(\mathbf{r},\hat{\mathbf{s}}) + (\hat{\mathbf{s}} \cdot \nabla + \mu_a(\mathbf{r}) + \mu_s(\mathbf{r}))\phi(\mathbf{r},\hat{\mathbf{s}}) - \mu_s(\mathbf{r}) \int \Theta(\hat{\mathbf{s}},\hat{\mathbf{s}}')\phi(\mathbf{r},\hat{\mathbf{s}}')d\hat{\mathbf{s}}' = 0, \tag{3}$$

where $\mathbf{r}$ denotes the position, $\hat{\mathbf{s}}$ and $\hat{\mathbf{s}}'$ denote unit direction vectors, and $\mu_a$ and $\mu_s$ denote the absorption and scattering coefficients of a medium, respectively. $q(\mathbf{r},\hat{\mathbf{s}})$ is the source contribution, $\phi(\mathbf{x},\hat{\mathbf{s}}')$ is the light radiance, and $\Theta(\hat{\mathbf{s}},\hat{\mathbf{s}}')$ is the scattering function representing the probability that the propagation direction is converted from $\hat{\mathbf{s}}$ into $\hat{\mathbf{s}}'$ if scattered. The RTE is derived from the conservation of energy, where optical properties including polarization, coherence and ionization are neglected [11].

The RTE can be simplified to the diffusion approximation provided that the scattering is high ($\mu_s \gg \mu_a$) and the scattering medium is nearly isotropic. Specifically, the radiance can be expanded as the combination of spherical harmonics where the basis sets consist of Legendre polynomials $P_n$ [12]. A truncated form using only the first-order set is sufficient under the scattering assumptions, and the resultant $P_1$ approximation is the diffusion approximation expressed as

$$Q(\mathbf{r}) = \mu_a(\mathbf{r})\Phi(\mathbf{r}) - \nabla \cdot [D(\mathbf{r})\nabla\Phi(\mathbf{r})], \tag{4}$$

where $Q(\mathbf{r}) = \int q(\mathbf{r},\hat{\mathbf{s}})d\hat{\mathbf{s}}$ is the source and $\Phi(\mathbf{r}) = \int \phi(\mathbf{r},\hat{\mathbf{s}})d\hat{\mathbf{s}}$ is the fluence. The parameter D is the diffusion coefficient given as

$$D = \frac{1}{3\mu_s'}, \quad \mu_s' = \mu_s(1-g) = \mu_s\left(1 - \int (\hat{\mathbf{s}}' \cdot \hat{\mathbf{s}})\Theta(\hat{\mathbf{s}},\hat{\mathbf{s}}')d\hat{\mathbf{s}}'\right), \tag{5}$$

where $\mu_s'$ is the reduced scattering coefficient and $g$ is the anisotropy factor. For a homogeneous medium ($D(\mathbf{r}) = D$, $\mu_a(\mathbf{r}) = \mu_a$, $\mu_s'(\mathbf{r}) = \mu_s'$) and a point source ($Q(\mathbf{r}) = \delta(\mathbf{r} - \mathbf{r}')$), the solution to the diffusion equation using the Green's function is given as

$$\Phi(\mathbf{r}) = \frac{1}{4\pi D|\mathbf{r}-\mathbf{r}'|}\exp(-\mu_{eff}|\mathbf{r} - \mathbf{r}'|), \tag{6}$$

where $\mu_{eff} = \sqrt{3\mu_a\mu_s'}$ [12-14] is the effective attenuation coefficient and $\mathbf{r}'$ is the spatial position of the source.

The solution can be modified for our acquisition environment in which a pencil beam is obliquely incident on a semi-infinite scattering medium. As illustrated in Fig. 2, the beam can be converted into two isotropic point sources mirror-symmetric about the extrapolated boundary $z = z_b$ due to the refractive-index mismatch between media. Note that every fiber is mounted on the transducer and covered by a thick acrylic holder and protection glass (BK-7 optical glass). Thus, the main contributor to the mismatch with the scattering medium is the solid transducer rather than the ambient medium (air). Appendix A1 describes the detailed computation of $z_b$ using the refractive indexes $n$ of the media.

The relative positions of the point sources with respect to the boundary are determined by the transport mean free path $l_t = \frac{1}{\mu_s'}$. For the kth fiber whose tip position is $\mathbf{r}_k' = (x_k', y_k', z_k')$ and incident angle $\theta$, the fluence solution can be expressed as

$$\Phi_k(\mathbf{r}) = \alpha_1 \left(\frac{\exp(-\mu_{eff}|\mathbf{r}-\mathbf{r}_k^+|)}{4\pi D|\mathbf{r}-\mathbf{r}_k^+|} - \frac{\exp(-\mu_{eff}|\mathbf{r}-\mathbf{r}_k^-|)}{4\pi D|\mathbf{r}-\mathbf{r}_k^-|}\right), \tag{7}$$

where $\alpha_1$ is a scalar, $\mathbf{r}_k^+ = (x_k', y_k' - l_t\sin\theta, z_k' + l_t\cos\theta)$ and $\mathbf{r}_k^- = (x_k', y_k' - l_t\sin\theta, -z_k' - l_t\cos\theta - 2z_b)$. Note that the unknown parameters are $\alpha_1$, $\mu_s'$ and $\mu_{eff}$. Other parameters can be derived from them or known initially.

If the positions of two imaginary isotropic sources in Fig.2 are close to each other due to a very small $l_t$ (e.g. very high $\mu'_s$), the fluence can be simplified to an asymptotic expression

$$\Phi_k(\mathbf{r}) = \alpha_2 \frac{z_k(1+\mu_{eff}|\mathbf{r}-\mathbf{r}_k|)}{|\mathbf{r}-\mathbf{r}_k|^3} \exp(-\mu_{eff}|\mathbf{r} - \mathbf{r}_k|), \tag{8}$$

where $\alpha_2$ is a scalar. The derivation is presented in Appendix (B). Here, unknown parameters are only $\alpha_2$ and $\mu_{eff}$, i.e. $\mu_{eff}$ is the only parameter defining the relative distribution of laser fluence in the medium. Held et al. employed this model [3].

In this paper, analytic fluence expressions in Eqs. (7) and (8) are called Model I and II, respectively. Thus, when Model II is quite accurate, it makes fluence estimation from experimental data very simple and stable. We will explore below how well both models compare to Monte-Carlo simulations, and under what conditions these models can be used for laser fluence assessment from swept-beam PA measurements.

*2.3 Optical Fluence Estimation*

PA measurement can be expressed as

$$y_{j,k}(\mathbf{r}_i) = p_{j,k}(\mathbf{r}_i) + n_{j,k}(\mathbf{r}_i), \tag{9}$$

where $n_{j,k}(\mathbf{r}_i)$ is system noise, and $j$ and $k$ are wavelength index and optical fiber index, respectively. Control data were recorded by assigning zero laser power at the first wavelength ($j = 0$) to estimate the noise bias as $b_k = \frac{1}{|\Omega|}\sum_{i\in\Omega} y_{1,k}(\mathbf{r}_i)$. The unbiased measurement can then be obtained as

$$\bar{y}_{j,k}(\mathbf{r}_i) = y_{j,k}(\mathbf{r}_i) - b_k = p_{j,k}(\mathbf{r}_i) + \bar{n}_{j,k}(\mathbf{r}_i), \qquad j \in \{1,2,\ldots,9\}, \tag{10}$$

where $E(\bar{n}) = 0$. Note that the unknown parameters needed to estimate $p_{j,k}(\mathbf{r}_i)$ are $\mu_s^{'(j)}$, $\mu_{eff}^{(j)}$, and $\beta_{i,j} = \Gamma\alpha_1\bar{\mu}_a^{(j)}(\mathbf{r}_i)$ provided Model I is employed. To enhance estimation efficiency, we selected position indices as $\tilde{\Omega} = \{i | \sum_{j,k} \bar{y}_{j,k}(\mathbf{r}_i) > \tau\}$ where $\tau$ denotes the threshold value. Also, we normalized measurements to reduce the degrees of freedom as

$$\tilde{y}_{j,k}(\mathbf{r}_i) = \frac{\bar{y}_{j,k}(\mathbf{r}_i)}{\sum_{k=1}^{20} \bar{y}_{j,k}(\mathbf{r}_i)} \approx \frac{\Phi_{j,k}(\mathbf{r}_i)}{\sum_{k=1}^{20} \phi_{j,k}(\mathbf{r}_i)} + \tilde{n}_{j,k}(\mathbf{r}_i), \qquad i \in \tilde{\Omega}, \tag{11}$$

where $\tilde{y}_{j,k}(\mathbf{r}_i)$ now depends on two parameters, $\mu_s^{'(j)}$ and $\mu_{eff}^{(j)}$ for every $i$, $j$ and $k$.

The normalized version of fluence $\widetilde{\Phi}_{j,k}(\mathbf{r}_i) = \frac{\Phi_{j,k}(\mathbf{r}_i)}{\sum_{k=1}^{20} \phi_{j,k}(\mathbf{r}_i)}$ ranges from 0 to 1. The optimal parameters for the $j$th wavelength can be estimated as

$$\left(\hat{\mu}_{eff}^{(j)}, \hat{\mu}_{s}^{\prime(j)}\right) = \operatorname*{argmin}_{\left(\mu_{eff}^{(j)}, \mu_{s}^{\prime(j)}\right)} \sum_{\mathbf{r}_i \in \Omega} w_i \sum_{k=1}^{20} \left| \widetilde{y}_{j,k}(\mathbf{r}_i) - \frac{\phi_{jk}(\mathbf{r}_i)}{\sum_{k=1}^{20} \phi_{jk}(\mathbf{r}_i)} \right|^2 \quad (12)$$

where $w_i = \sum_{j,k} \bar{y}_{j,k}(\mathbf{r}_i)$ is the weight such that a position with higher SNR contributes more to the estimate. Likewise, Model II can be used for estimation, but the search parameter is only $\hat{\mu}_{eff}^{(j)}$ for every wavelength index $j$. The fluence estimate $\widehat{\Phi}_{j,k}(\mathbf{r}_i)$ can be obtained by substituting optical parameter estimates into either Model I or II.

*2.4 Fluence Correction of Light Absorption Spectrum*

Assume that one type of chromophore is located at a local position of interest and its absorption spectrum is known as $\alpha_j$. The optical absorption spectrum obtained from PA measurements $d_j = \sum_k \bar{y}_{j,k}$ is distorted due to the wavelength-dependent fluence of the surrounding turbid medium (biological tissue, for example), as shown in Eq. (1). The spectrum can be corrected using the fluence estimate $\widehat{\Phi}_{j,k}$ as $c_j = \frac{\sum_k \widehat{\Phi}_{jk} \bar{y}_{j,k}}{\sum_k \widehat{\Phi}_{jk}^2}$. By comparing $d_j$ and $c_j$ with $a_j$, fluence estimation accuracy can be computed.

## 3   Numerical Simulations

*3.1. Simulation parameters*

The primary purpose of simulations is to verify the proposed fluence models and associated estimation methods described in Sections 2.2 and 2.3. First, Models I and II were compared to ground truth Monte Carlo simulations. We adopted the medium geometry of Fig. (2) for all simulations. All media were approximately represented as cubical shapes. Their sizes, optical parameters, and refractive indexes are summarized in Table 1. We assumed that one of the fibers transmits light into the scattering medium at the interface. The location and tilt angle of the fiber tip are (0 mm, 5.68 mm, 0 mm) and 35°, respectively. Total photon number for the Monte Carlo tests is 20 million. The simulation used the open source program, MCX Studio [15].

**Table 1. Range of parameters in numerical simulations**

|  | Cuboid size ($x$ mm × $y$ mm × $z$ mm) | $\mu_a$ (cm$^{-1}$) | $\mu_s'$ (cm$^{-1}$) | Refractive index $n$ |
|---|---|---|---|---|
| Ambient medium | (50 × 50 × 50) | 0.0 | 0.00 | 1.00 |
| Scattering medium | (50 × 50 × 50) | 0.01 − 0.05 | 5 − 35 | 1.33 |
| Transducer medium | (30 × 20 × 30) | 1000.0 | 1000.0 | 1.49 |

We also compared errors in parameter estimation for Model I compared to Model II at a fixed optical wavelength. We assumed a point target located at position $\{\mathbf{r} = (x \text{ mm}, 0 \text{ mm}, z \text{ mm})\}$ in a scattering medium whose optical parameters are $\mu_a$, $\mu_s'$ and $\mu_{eff}$. For every $(z, \mu_s')$, we conducted 100 simulation tests to estimate $\mu_{eff}$ by randomly changing values of other parameters in the range. A fractional error sample for every test is given as $\epsilon = (\mu_{eff} - \hat{\mu}_{eff})/\mu_{eff} \times 100$ (%), where $\hat{\mu}_{eff}$ denotes the estimate using Model I. We averaged 100 error samples for every $(z, \mu_s')$ to monitor the mean error pattern $\bar{\epsilon}(z, \mu_s')$ over penetration depth and scattering coefficient.

In the last simulation, we investigated the accuracy of parameter estimation for different wavelengths in a turbid medium where $\mu_a$ and $\mu_s'$ are wavelength dependent. We generated synthetic data using Models I and II and added white Gaussian noise to mimic experimental conditions and explore reconstruction algorithm stability to noise. The total number of SNR levels was 15, ranging from 20 dB to 50 dB, and the number of test data sets for each SNR was 100. SNR is defined as

$$\text{SNR} = 10 \log_{10}\left(\left(\frac{1}{20}\sum_{k=1}^{20} p_k\right)^2 / \sigma_n^2\right), \tag{13}$$

where $\sigma_n^2$ is the noise variance. A point target is assumed to be located at position $\{\mathbf{r}_i = (0, 0, 10 \text{ mm})\}$ in a scattering medium whose optical parameters are $\mu_s'^{(j)}$, $\mu_a^{(j)}$ and $\mu_{eff}^{(j)}$ at the $j$th wavelength.

According to the literature [7], the scattering coefficient in brain varies the most over wavelength of all tissue types [14, 15]

$$\mu_s'(\lambda) = 40.8(\lambda/500 \text{ nm})^{-3.089}, \tag{14}$$

providing the most complicated conditions for light absorption spectrum reconstruction using PA signals. We used this scattering function for the test medium in the third simulation but assumed the absorption coefficient $\mu_a^{(j)} (= 0.03 \text{ cm}^{-1})$ is constant for all wavelengths. This does not reduce the generality of the results because (as we show below) reconstruction error is not highly sensitive to variations in light absorption. Table 2 summarizes parameter values for $j \in \{1,2,...,8,9\}$.

Optical parameters were estimated from noise contaminated data using Eq. (12) based on either Model I or II. Then, the fluence estimate $\widehat{\Phi}_{j,k}(\mathbf{r}_i)$ was compared with ground truth $\Phi_{j,k}(\mathbf{r}_i)$ using the correlation coefficient as

$$\rho = \frac{\sum_k (\hat{\phi}_k \phi_k)}{\sqrt{\sum_k \hat{\phi}_k^2 \sum_k \phi_k^2}}, \tag{15}$$

where $\hat{\phi}_k = \sum_{i,j} \widehat{\Phi}_{j,k}(\mathbf{r}_i) - \frac{1}{20}\sum_{i,j,k} \widehat{\Phi}_{j,k}(\mathbf{r}_i)$ and $\phi_k = \sum_{i,j} \Phi_{j,k}(\mathbf{r}_i) - \frac{1}{20}\sum_{i,j,k} \Phi_{j,k}(\mathbf{r}_i)$. We examined the mean correlation coefficient using 100 estimation samples for every SNR level.

**Table 2. Optical parameters for the light fluence assessment with noise contaminated data**

|  | value |
|---|---|
| $\mu_s'^{(j)} (cm^{-1})$ | [13.53, 12.42, 11.43, 10.55, 9.75, 9.03, 8.38, 7.79, 7.25] |
| $\mu_{eff}^{(j)} (cm^{-1})$ | [1.10, 1.05, 1.01, 0.97, 0.93, 0.90, 0.86, 0.83, 0.80] |

*3.2. Simulation Results*

Fig. 3a shows slices of a 3D fluence distribution $\Phi(x, y, z)$ from the Monte Carlo simulation. We used the distribution of optical fluence $\Phi(0,0,z)$ along the **z** axis as a reference to compare with corresponding distributions calculated using Model I and Model II (given by eqs. (7) and (8) respectively). Figures 3c-e show the distributions obtained for different $\mu_s'$ (2 cm$^{-1}$, 5 cm$^{-1}$ and 10 cm$^{-1}$, respectively). The elevational position $y'$ of the source is fixed at 5.70 mm, which corresponds to the fiber positions in our experimental transducer array (see Fig.2). Under the condition of extremely small $\mu_s' (= 2 \text{ cm}^{-1})$, Monte Carlo produces a spike since ballistic transport crosses the image plan. As expected, Model I is very close to Monte Carlo results for diffusive conditions because scattering is dominant in the medium. Model II approximates Model I when the measurement point is located at distances from the source and the interface much higher than the transport mean free path $l_t$. The higher the scattering, the closer the agreement between Model II and Monte Carlo, as shown in Figs. 3d,e. We also compared fluence models at different lateral source positions ($y'$ is 2.85 mm, 5.70 mm and 11.40 mm, see Figs. 3f-h). Scattering $\mu_s'$ was set to 10 cm$^{-1}$.

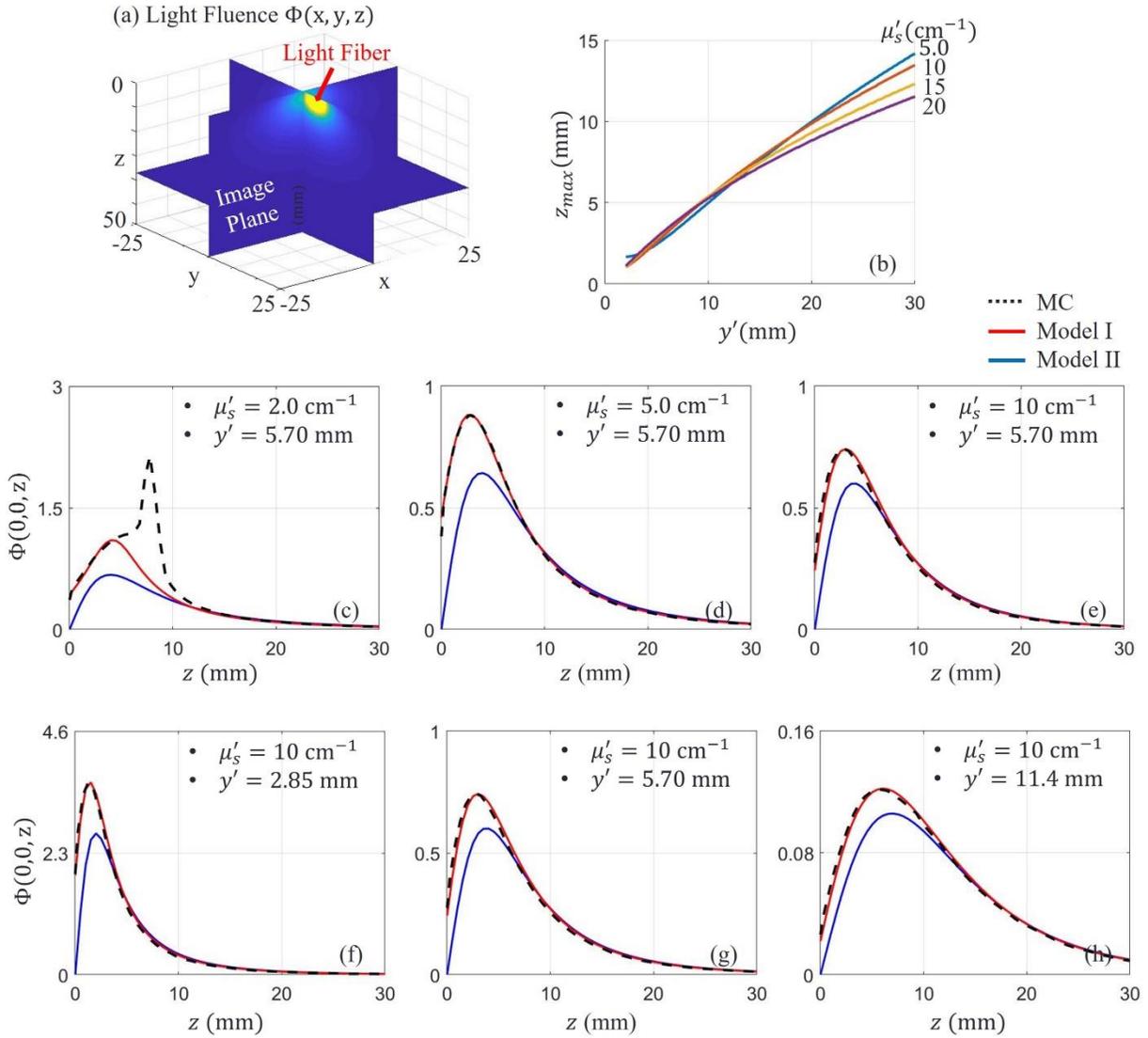

**Fig. 3** (a) Slices (planes) of 3D light fluence $\Phi(x, y, z)$ in a scattering medium. The magnitude distribution is obtained by Monte Carlo (MC) simulation. (b) Position (axial) $z_{max}$ of the maximum optical fluence location along line $(0,0,z)$ when a fiber source is located at different lateral positions $(0, y', 0)$. The graphs represent $z_{max}$ over $y'$ for several scattering (e.g. different $\mu_s'$) conditions. (c-e) Distributions of optical fluence for different $\mu_s'$, but fixed source position $y'$. (f-h) Distributions of optical fluence for different source position $y'$ but fixed $\mu_s'$. The light absorption coefficient of the medium is $\mu_a = 0.03$ cm$^{-1}$.

The position $z_{max}$ of maximum optical fluence along the **z** axis increases with increasing distance $y'$ between the imaging plane and light source, and Model II converges quickly to Model

I soon after $z_{max}$. Figure 3b summarizes the shift in $z_{max}$ as a function of source position $y'$ for different $\mu_s'$ typical of biological tissues [7]. For the source position in our system ($y' = 5.7$ mm), $z_{max}$ barely changes for $\mu_s'$ varying from 5-20 cm$^{-1}$. This means that simplified Model II converges to MC and Model I at very similar depths over a wide range of tissue scattering. For our transducer configuration, $z_{max}$ is around 3 mm. This fact is very important for practical fluence assessment because $\mu_s'$ is not known a priori.

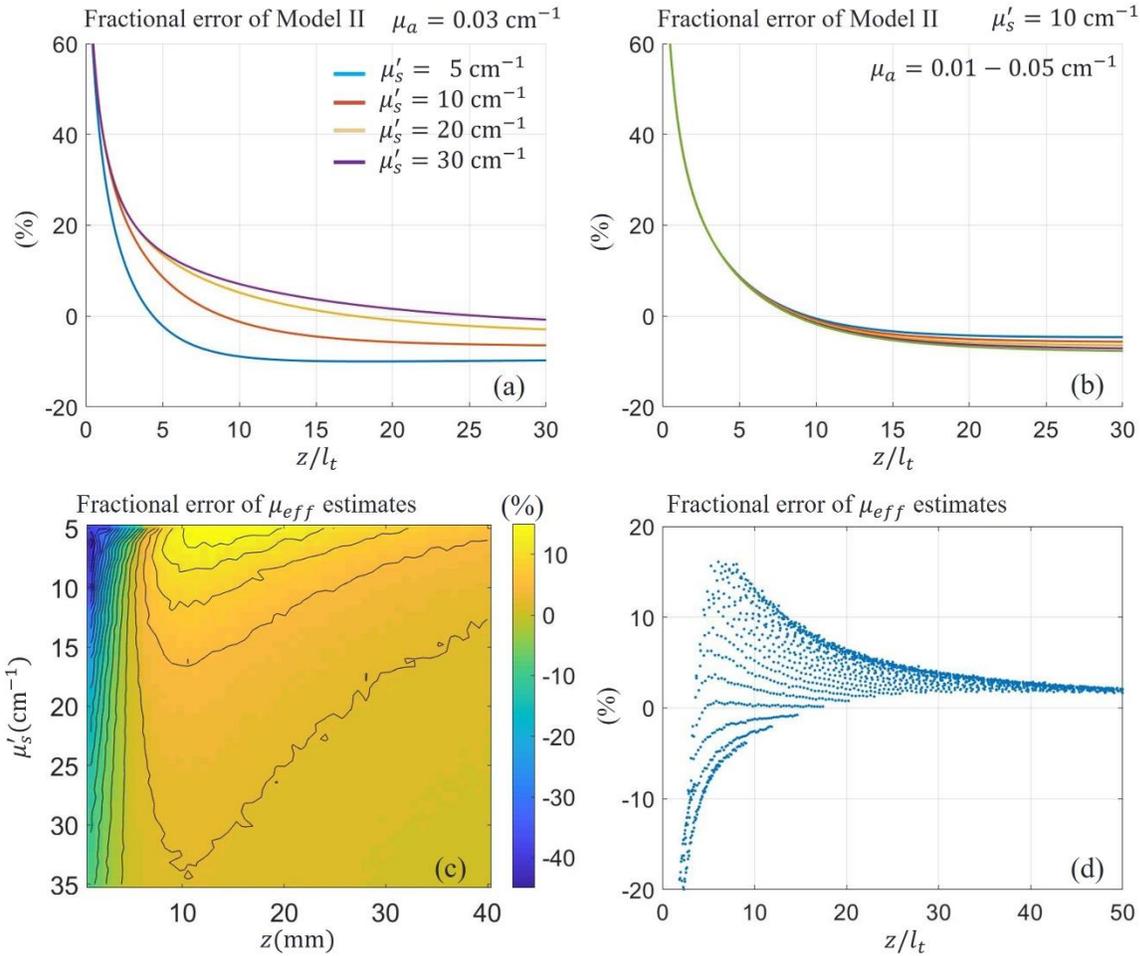

**Fig. 4** (a) Fractional errors (relative percentage changes) of Model II with respect to Model I over $z/l_t$ when the reduced scattering coefficient $\mu_s'$ is 5, 10, 20 and 30 cm$^{-1}$. The common light absorption coefficient $\mu_a$ is 0.03 cm$^{-1}$. (b) Fractional errors over $z/l_t$ when the $\mu_a$ value is 0.01, 0.02, 0.003, 0.04 and 0.05 cm$^{-1}$. The common reduced scattering coefficient $\mu_s'$ is 10 cm$^{-1}$. (c) Mean fractional error $\bar{\epsilon}(z, \mu_s')$ of $\mu_{eff}$ estimates with respect to the ground truth over 2D domains, axial depth z, and reduced scattering coefficient $\mu_s'$. (d) Mean fractional error $\bar{\epsilon}(z, \mu_s')$ over 1D axis $z/l_t = z\mu_s'$. Pixel values of (c) are represented by dots on this graph.

Based on these initial simulations, it's clear that Model I can be used as ground-truth if the light source is located more than a distance $l_t$ from the light source. We now focus on the difference in parameter estimation error (bias) between Models 1 and II in more detail. The fractional error for Model II is defined as $(\Phi^{(I)} - \Phi^{(II)})/\Phi^{(I)} \times 100$ (%), where $\Phi^{(I)}$ and $\Phi^{(II)}$ denote Model I and Model II, respectively. Figure 4a shows the error over $z/l_t$ for several $\mu_s'$ over the tissue range. The error is under 10% when $z/l_t$ is larger than ~10. Figure 4b shows the error for several assumed $\mu_a$ values. Note that the $\mu_a$ variation rarely changes the error. Figure 4c illustrates the fractional error $\bar{\epsilon}(z, \mu_s')$ for $5 < \mu_s' < 35$ cm$^{-1}$ and $0 < z < 40$ mm when Model II is used to estimate $\mu_{eff}$. Figure 4d displays pixel values of (c) over the $z/l_t$ (= $z\mu_s'$) axis.

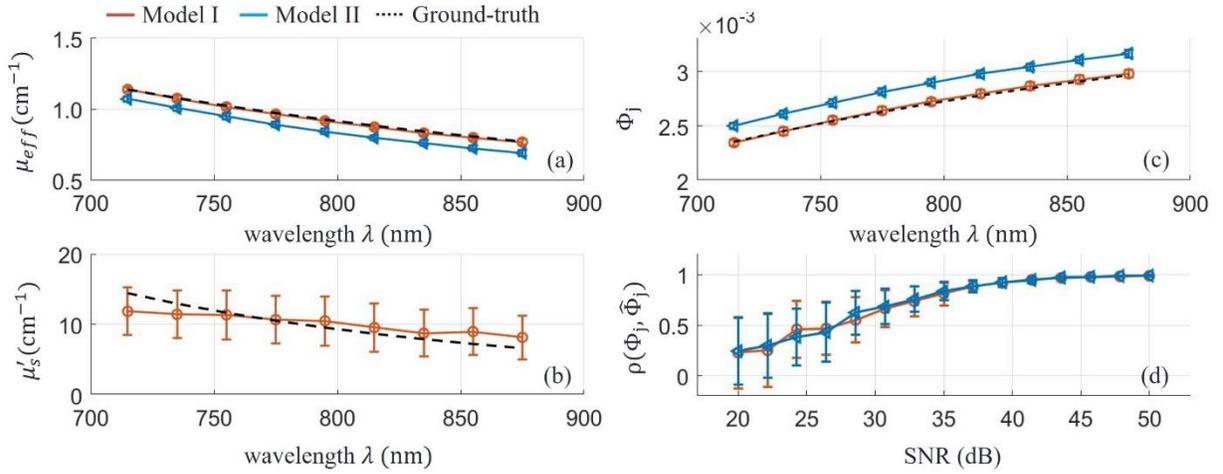

**Fig. 5** (a-c) Optical parameter estimates and fluence estimates over wavelength. The reduced scattering coefficient $\mu_s'$ and the effective attenuation coefficient $\mu_{eff}$ of the medium vary with wavelength. The measurement SNR is 50dB. (a) Shows estimates of $\hat{\mu}_{eff}$ using Model I and Model II, and (b) shows estimated $\hat{\mu}_s'$ using Model I. (c) Shows $\hat{\Phi}_k(\mathbf{r}_i)$ estimates at position $\mathbf{r}_i$ using Model I and Model II. The markers (o) and (◁) indicate estimates using Model I and II, respectively. The dotted line denotes ground truth. (d) Shows the strength of association between the fluence spectrum $\Phi_k(\mathbf{r}_i)$ and its estimate $\hat{\Phi}_k(\mathbf{r}_i)$ over the SNR range. The markers (o) and (◁) indicate the correlation coefficients using Model I and II, respectively.

Figure 5 presents optical parameter estimates using Models I and II when data are contaminated by noise. We used eq. (14) to set ground-truth scattering coefficient variations over wavelength (715-875 nm). Figures 5a-c show estimates when the SNR is the highest (50 dB). Figure 5d shows the correlation between the estimated optical fluence spectrum and ground truth over a wide range of SNR. Model I estimation error (bias and deviation) is only a function of measurement noise. Estimates of $\mu_{eff}$ are unbiased at high SNR while those of $\mu_s'$ are biased and their errors are high even at this high SNR level. However, despite large error in $\mu_s'$ estimated by Model I, the correlation between resultant fluence estimates and ground truth is close to 1. As shown in Fig. 5d, the lower SNR level causes lower correlation because both estimation deviation and bias are higher. In Model II, estimation error is caused by model discrepancy as well as measurement noise. Note that Model II underestimates $\mu_{eff}$ at high SNR, but the ratio of estimate $\hat{\mu}_{eff}$ to ground truth $\mu_{eff}$ remains almost constant over the wavelength range. Thus, it provides competitive correlation between fluence estimate and ground truth despite the bias. The lower SNR, of course, leads to lower estimation performance (see Fig.5d).

## 4. Experiments

Phantom experiments produce realistic PA measurements to help validate the proposed methods. We conducted two studies.

*4.1. Phantom Study I*

In the first study, we explored the accuracy of optical fluence estimates in a turbid medium using a human hair as the absorbing target. Each hair was aligned parallel to the transducer elevational direction (y-axis) so that it appears as a point target in the x-y image plane (see Fig. 6a). The target was positioned in a cubic tank (open on top) filled with an optical scattering medium, and a transducer was positioned at the medium surface. The tank was sufficiently large so that boundary effects from all faces except the top interface between media can be neglected. The optical medium was an intralipid solution (20% IV fat emulsion, Fresenius Kabi, Deerfield, USA) with homogeneous scattering. We diluted the original 20% emulsion to 0.5%-4% ones to control the scattering coefficient.

Radio frequency ultrasound data were recorded using the Verasonics system and a 128 element linear-array transducer (LA 15/128-1633, Vermon S.A. Tours, France). The transducer center frequency is 15 MHz and the 3-dB bandwidth is 11-19 MHz. The array element pitch is 0.1 mm. The laser pulse energy was between 0.4 mJ and 0.5 mJ at the tissue surface, depending on the wavelength. A little (4% portion of it) was taken from the main path by a beam splitter and then recorder for every laser pulse by photodetector located before the fiber bundle. Thus, the energy of every laser pulse was measured to compensate its dependence on wavelength and take into account pulse-to-pulse energy variations (~ 8%). The number of axial samples and scan-lines were adjusted to the field of view (25.2 mm × 12.7 mm). One PA data set consists of 2048 samples × 128 channels × 20 fibers × 10 wavelengths × 40 frames acquired at a 1 kHz repetition rate, representing a total acquisition period of 8 sec. We averaged every data set over frames to enhance SNR and processed PA signals using conventional delay-and-sum beamforming to obtain an image (512 × 128) for every fiber and wavelength. Then, a Hilbert transform was applied to obtain the smooth envelope of axial image samples. The enveloped image was used as $y_{j,k}$ in Eq. (9) for the $k$th fiber and $j$th wavelength.

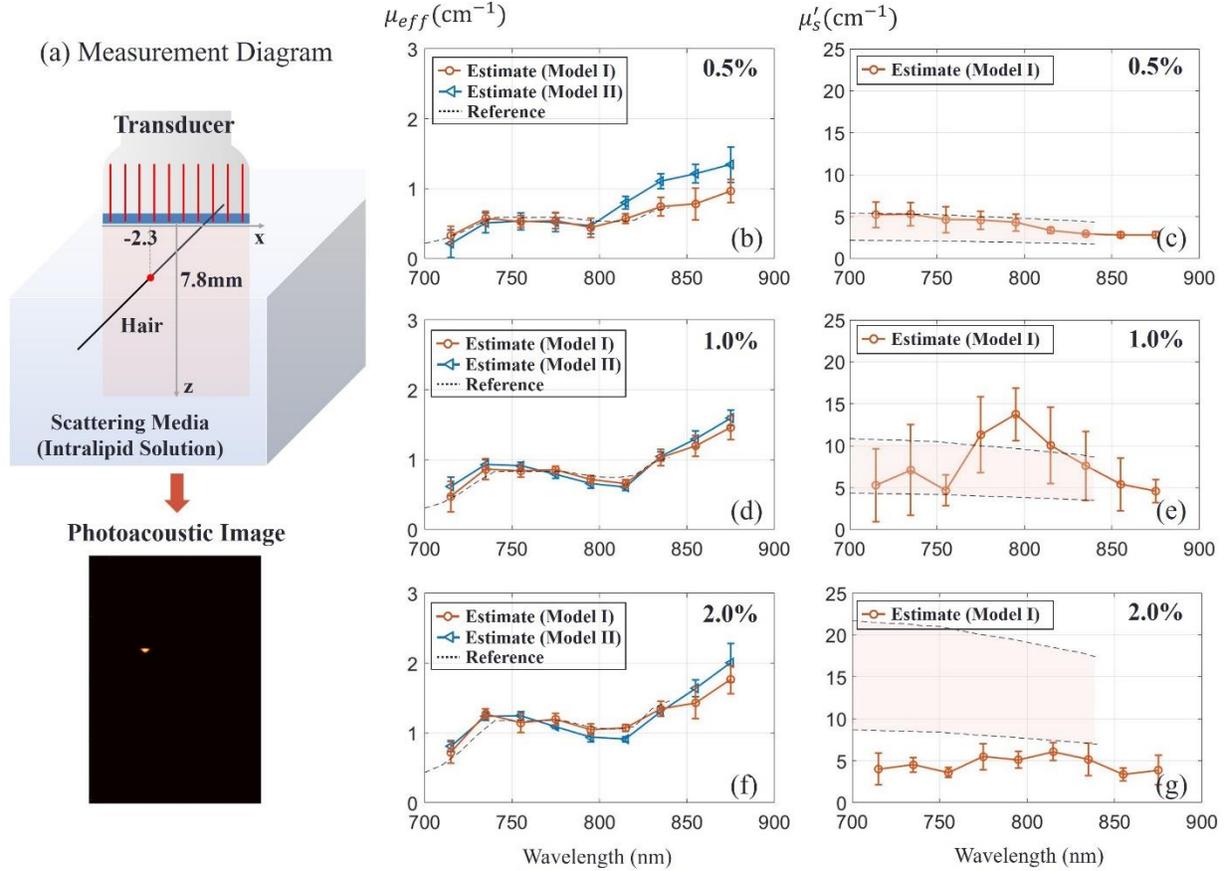

**Fig. 6**. Experimental assessment of turbid medium optical properties using the fast-swept concept. (a) Measurement diagram with a human hair as absorbing target immersed in a turbid medium (intralipid solution of different concentrations as indicated in top right corners of panels). Optical properties (reduced scattering coefficient $\mu_s'$ and effective attenuation coefficient $\mu_{eff}$) of the medium were determined using Model I (orange) and Model II (blue - $\mu_{eff}$ only). The point and error bar indicate estimation mean and standard deviation. A total of 8 samples (8 datasets) were used for statistics. The dotted line ($\mu_{eff}$) and the shaded region between dotted lines ($\mu_s'$) indicate the estimated value and range, respectively, for the results of Ref. [3].

Figure 6b illustrates optical parameter estimates over the range of laser wavelengths (715-875nm) for the 0.5% intralipid solution as the scattering medium. Each marker and error bar indicate mean and standard deviation of 8 samples, respectively. The reference spectrum employed here was obtained from Ref. [3]. Since this spectrum was measured for a 1.2% intralipid concentration in [3], it was scaled here to compare with the 0.5% solution and plotted as dotted lines (Fig. 6c). The expected reduced scattering coefficient shown on the right is presented as a shaded region since reference papers [3, 16-18] give different measurement values. The scattering coefficient $\mu_s'$ is under 5 cm$^{-1}$ over the measured wavelength range, close to the lower edge of medium scattering used in the numerical simulations. As shown in Figs. 6b,c, estimated $\mu_s'$, and $\mu_{eff}$ match very closely with those reported in the literature under similar measurement conditions. Due to the small $\mu_s'$, Model II provides a less accurate spectral estimate.

Estimates of effective attenuation coefficient for three higher intralipid concentrations (0.5%, 1% and 2%) are presented in Figs. 6b, 6d and 6f respectively. As expected, both Models I and II provide very accurate assessment of $\mu_{eff}$, whereas $\mu_s'$ estimates from Model I are clearly not accurate for these higher scattering media (Figs. 6e,g). However, as shown in the numerical simulations, one parameter, $\mu_{eff}$, is enough when Model II is valid. Thus, inaccuracies in the reconstructed $\mu_s'$ do not significantly affect the accuracy of laser fluence estimates.

## 4.2. Phantom Study II

The purpose of second study is to (i) perform optical fluence estimation in the medium, and (ii) apply optical fluence corrections to spectroscopic PA measurements. For this test, the container held three cylindrical tubes aligned parallel to the y-axis (see Fig.7a). We injected a nanoparticle solution (gold nanorods (GNR), width (11.4 nm), length (44.8 nm), mass concentration (2.2 mg/ml), longitudinal peak (776 nm), NanoHybrids Inc. Austin, USA) [10] and a black ink solution (Higgines Black Magic Ink, Chartpark Inc., Leeds, USA) as absorbers in Tubes I and III, respectively, where the absorption spectra of the solutions are well-known. Also, we injected water into Tube II as a control. The container was filled with a 1% intralipid solution as the scattering medium. We additionally added customized Prussian blue ink [10] to the solution to increase spectral distortion at the expense of high attenuation (low data SNR).

We used a total of 8 datasets and averaged them to improve SNR. Fig. 7 (a) presents the PA images at three particular wavelengths, where every pixel value is proportional to the photoacoustic signal magnitude presented on a log scale. Due to the limited view and bandwidth of the transducer, the signal only appears at the top and bottom of the tube. The signal beneath the tubes is from a reverberant wave in the tube. The region near Tube II has a weak signal because the tube material weakly absorbs light over this wavelength range. The ranges of SNR in Tube I and Tube III are 27.8-39.9 dB and 41.1-47.5 dB, respectively, over the wavelength range.

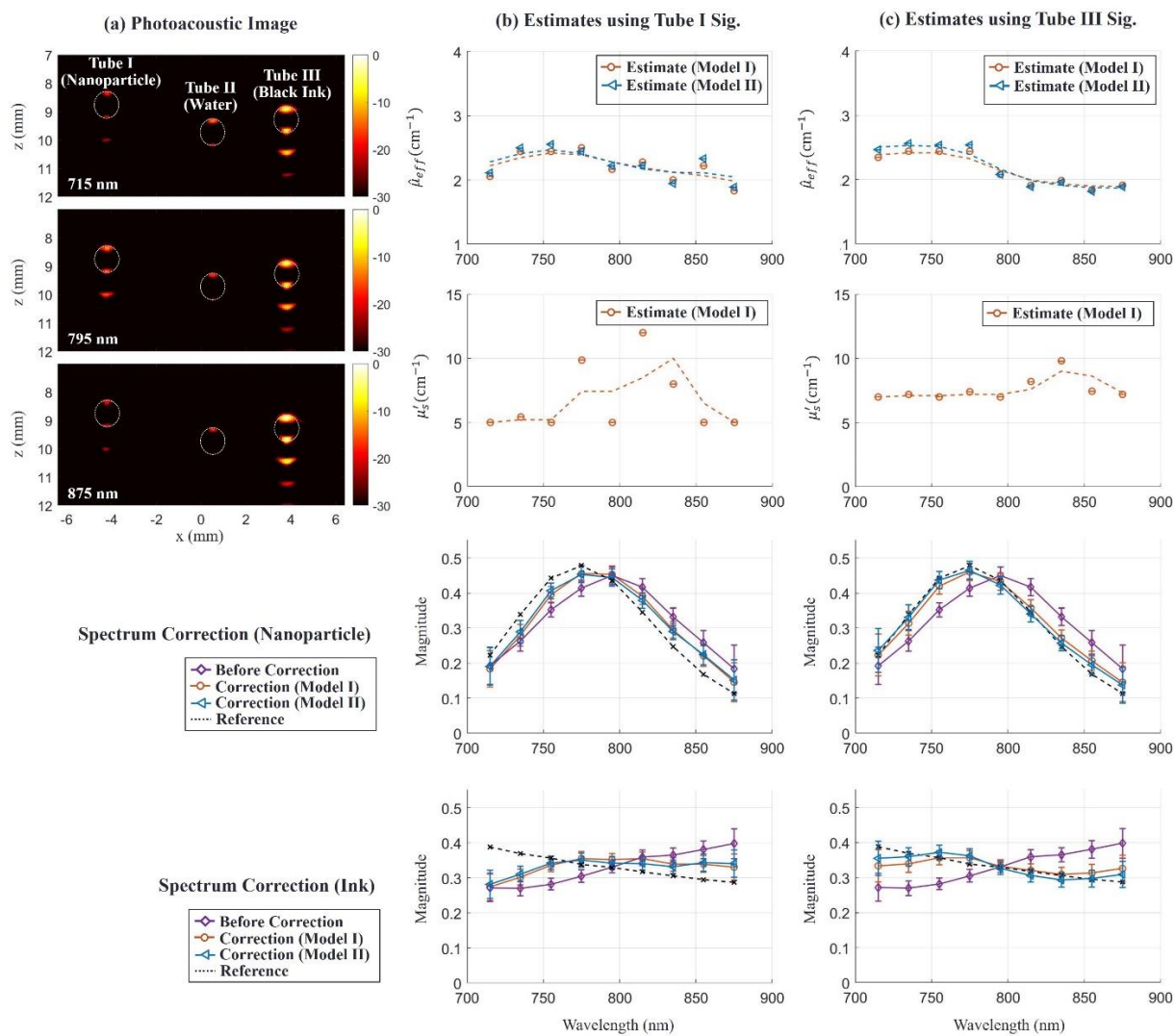

**Fig. 7**. Spectroscopic PA imaging of absorbing targets in a turbid medium. (a) Examples of PA images displayed over a log (dB) scale at 3 different wavelengths (715 nm, 795 nm and 875 nm) for Phantom Study II. The first, second and third tubes were filled with GNR solution, water and black ink, respectively. The scattering medium was a solution of intralipid and Prussian blue ink. The pixel value of the image is associated with the PA pressure. Dotted circles in the image indicate tube cross-sections. (b) and (c) columns show parameter estimates using nanoparticle signals in Tube I and black signals in Tube III, respectively, and spectrum correction results using the estimates. First and second rows represent estimates of effective light attenuation and reduced scattering coefficients in a turbid medium, respectively. Points marked (o) and (◁) denote estimates using Model I and Model II, respectively. The dotted line is obtained by smoothing the estimates over wavelength. Third and fourth rows show absorption spectra of the nanoparticle and black ink, respectively. Magenta curves with points marked (◇) correspond to measured PA absorption spectra, i.e. without fluence correction. Orange and blue curves with points marked (o) and (◁) respectively correspond to ink and GNR absorption spectra obtained after applying fluence correction using Model I and Model II, respectively. Dotted lines indicate the reference target spectra measured independently with optical spectrophotometry.

Fig. 7 (b) and (c) show estimation results using signals in Tube I and Tube III for laser fluence assessment, respectively. In other words, we demonstrate here that any target, independent of its absorption properties (for example, for a solution GNR or black ink), effectively absorbing light over the spectral range probed by the system can be used for the laser fluence estimation in the swept-beam concept. In practice, such targets can be blood vessels, injected contrast agents or labelled drugs, or other absorbing targets.

The first and second row in columns (b) and (c) of Fig.7 show effective light attenuation and reduced scattering coefficient estimates, respectively. We smoothed all estimates over wavelength to reduce bias and applied them for fluence compensation. Note that the signals in Tube III provide more stable estimates than those in Tube I due to higher SNR, as predicted in numerical simulations above. The third and fourth row compare raw PA spectra and spectra after fluence correction for nanoparticles and black ink, respectively. Corrected spectra closely match ground truth. As predicted in simulations, high SNR increases correction accuracy. Note that laser fluence estimation was performed here using only PA signals from a single target, yet reconstruction accuracy is already reasonable. If a larger number of points within the PA image are used simultaneously for fluence correction, then more accurate corrections can be obtained.

## 5. Discussion and Conclusions

In simulations, we investigated optical parameter estimation using measurements on chromophores acting as targets in a turbid medium. Estimation results clearly depend on the number, position, and size of targets. More and bigger targets produce lower estimation bias and variance at any SNR. Also, targets located near image edges contribute more to estimation results than those near the center because they provide a wider range of distances $|\mathbf{r} - \mathbf{r}'_k|$ between target and source positions. Except for targets located near the image center line, estimation accuracy is maintained relatively independent of the distribution of absorption targets for a given SNR.

The primary advantage of Model I is that it more closely matches the result of MC simulations when the source-target distance exceeds the photon transport mean free path. Thus, estimates of $\mu_{eff}$ based on this model are unbiased and accurate at zero noise levels, as shown in Fig. 6. A disadvantage of this model is the additional parameter, $\mu'_s$, that must be simultaneously estimated. A two parameter search often produces higher estimation errors under real, not ideal, conditions, especially when experimental measurements are contaminated by noise. Thus, using constraints for narrowing the search range of $\mu'_s$ based on prior knowledge would increase the performance under low SNR conditions.

Model II is an approximate version of Model I where $\mu'_s$ can be considered infinitely high. In reality, this approximation works when a target is located far (i.e. at distances much larger than $l_t$) from all light sources and interfaces. As shown in Figs. 3 and 4, within the range of typical light scattering of biological tissues, Model II can be used for targets located at depths deeper than 5-7 mm from the medium surface, i.e. at depths where the wavelength dependence of laser fluence starts affecting spectroscopic measurements. As depth z increases, Model II converges to Model I very quickly. Note, that the ratio of $\hat{\mu}_{eff}$ to $\mu_{eff}$ remains mostly constant as $\mu_{eff}$ varies over wavelength, as shown in Fig. 5, even though we used the highest variation of $\mu'_s$ (for brain tissue) from all biological tissues reported in the literature. Model II is a single parameter model which, in the range where it is valid, provides fast and stable laser fluence estimates.

Phantom studies validated the overall performance of spectral corrections based on optical parameter estimation. Fluence-corrected spectra closely match ground-truth spectra, as shown in Figs. 6 and 7. Estimation variance depends on SNR at each wavelength. Thus, estimates based on the GNR target are less accurate because of relatively small signal amplitude at the edges (715nm and 875nm wavelengths) of the GNA absorption spectrum.

One of the assumptions in our models is that the laser fluence at one target is not shadowed by another target. Shadowing can be neglected if chromophores are located sparsely over the imaging field. This allows us to simply estimate the fluence distribution $\Phi(\mathbf{r})$ and then correct measured PA spectra.

In conclusion, our study shows that one of the fundamental problems of PA imaging, i.e. decoupling the local light absorption coefficient from the local optical fluence, may be solved automatically using a swept-beam imaging concept when the medium under study is irradiated sequentially with a narrow laser beam. Furthermore, in [10] we clearly demonstrated that a swept-beam geometry can be implemented with a high rep-rate, fast wavelength-tunable laser source scanned over a collection of optical fibers distributed around an ultrasound imaging array to produce 50 Hz interleaved spectroscopic PA and US images. We adopted an analytic laser fluence model for a homogenous scattering medium and estimated wavelength-dependent fluence variations within the image plane from PA measurements without knowledge of optical properties and without large computational costs. In phantom studies, we showed that fluence correction considerably improves the accuracy of measured spectra and can lead to quantitative estimation of relative chromophore concentrations. Real-time PAUS systems providing fluence-compensated spectroscopic PA imaging have the potential to enable molecular imaging for many clinical applications, such as interventional procedure guidance using molecularly labelled therapeutic agents and procedure validation based on spectroscopic confirmation of modifications in microvascular networks.

**Appendix**

*A.1 Refractive-Index-Mismatched Boundary*

Assume that the boundary between medium I and II is refractive-index-mismatched and one source to medium I is located at the interface. Outgoing light toward II from I is partially reflected at the boundary due to the mismatch. This can be mathematically expressed as

$$\int_{\mathbf{s}\cdot\mathbf{n}>0} L(\mathbf{r},\mathbf{s})\mathbf{s}\cdot\mathbf{n}\,d\Omega = \int_{\mathbf{s}\cdot\mathbf{n}<0} R(\mathbf{s}\cdot\mathbf{n})L(\mathbf{r},\mathbf{s})\mathbf{s}\cdot\mathbf{n}\,d\Omega, \tag{16}$$

where $L$ and $R$ denote the radiance and Fresnel reflection, respectively. The vectors $\mathbf{r}$, $\mathbf{s}$ and $\mathbf{n}$ denote the position on the boundary, unit direction vector, and unit normal vector pointing toward medium I, respectively. $d\Omega$ denotes a differential solid angle element. Using a spherical harmonics expansion, the radiance can be approximated as

$$L(\mathbf{r},\mathbf{s}) = \frac{1}{4\pi}\Phi(\mathbf{r}) + \frac{3}{4\pi}\mathbf{J}(\mathbf{r})\cdot\mathbf{s}, \tag{17}$$

where J is the current density, or the energy flow per unit area. The Fresnel reflection is given as

$$R(\mathbf{s}\cdot\mathbf{n}) = \begin{cases} \frac{1}{2}\left[\left(\frac{n_r\cos\theta' - \cos\theta}{n_r\cos\theta' + \cos\theta}\right)^2 + \left(\frac{n_r\cos\theta - \cos\theta'}{n_r\cos\theta + \cos\theta'}\right)^2\right], & if\ 0 \leq \theta \leq \theta_c \\ 1, & otherwise \end{cases} \tag{18}$$

where $\theta = \cos^{-1}(\mathbf{s}\cdot\mathbf{n})$ is the angle of incidence, $\theta' = \sin^{-1}(n_r\sin\theta)$ is the angle of refraction, $\theta_c = \sin^{-1}(n_r^{-1})$ is the critical angle, and $n_r$ is the ratio of the refractive index of medium I to that of medium II. Substituting Eq. (17) into Eq. (16) yields

$$\frac{1}{4}\Phi(\mathbf{r}) - \frac{1}{2}\mathbf{J}(\mathbf{r})\cdot\mathbf{n} = \frac{1}{4}R_\Phi\Phi(\mathbf{r}) - \frac{1}{2}R_J\mathbf{J}(\mathbf{r})\cdot\mathbf{n}, \tag{19}$$

where $R_\Phi = \int_0^{\pi/2} 2\sin\theta \cos\theta\, R(\cos\theta)d\theta$ and $R_J = \int_0^{\pi/2} 3\sin\theta\, (\cos\theta)^2 R(\cos\theta)d\theta$ .

Substituting Fick's law, $J(\mathbf{r}) = -D\nabla\Phi(\mathbf{r})$, to Eq. (19) results in

$$\Phi(\mathbf{r}) - 2D\frac{1+R_J}{1-R_\Phi}\frac{\partial\Phi(\mathbf{r})}{\partial z} = 0, \tag{20}$$

A Taylor series expansion to first order leads to the fluence $\Phi(z)$ at $z = -z_b = -2D\frac{1+R_J}{1-R_\Phi}$ to be approximately zero. The face $z = -z_b$ is called the extrapolated boundary.

## A.2 Simplified Diffusion Equation

Let $\frac{\exp(-\mu_{eff}|\mathbf{r}-\mathbf{r}'_k|)}{|\mathbf{r}-\mathbf{r}'_k|}$ be $f(x,y,z)$ for convenience. If the two imaginary sources are close, Eq. (7) simplifies to

$$\Phi_k(\mathbf{r}) \approx \lim_{l_t \to 0} \frac{3f(x, y+l_t\sin\theta, z-l_t\cos\theta)}{4\pi l_t} - \frac{3f(x, y+l_t\sin\theta, z+l_t(\cos\theta+\kappa))}{4\pi l_t}$$

$$= \frac{3}{4\pi}\left[\frac{\partial f}{\partial y}\sin\theta - \frac{\partial f}{\partial z}\cos\theta\right] - \frac{3}{4\pi}\left[\frac{\partial f}{\partial y}\sin\theta + \frac{\partial f}{\partial z}(\cos\theta+\kappa)\right]$$

$$= \frac{-3}{4\pi}(2\cos\theta + \kappa)\frac{\partial f(x,y,z)}{\partial z}$$

$$= \frac{3}{4\pi}(2\cos\theta + \kappa)\frac{z_k(1+\mu_{eff}|\mathbf{r}-\mathbf{r}'_k|)}{|\mathbf{r}-\mathbf{r}'_k|^3}\exp(-\mu_{eff}|\mathbf{r}-\mathbf{r}'_k|)$$

where $l_t = 3D$ and $\kappa = \frac{4(1+R_J)}{3(1+R_\Phi)}$. In Eq. (9), the constant $\alpha_2$ is $\frac{3}{4\pi}(2\cos\theta + \kappa)$.

*Acknowledgments*

We would like to greatly thank Professor Martin Frenz at University of Bern for several helpful discussions about the general principles of laser fluence estimation. This work was partially supported by NIH grant HL-125339, GE Healthcare, and the Department of Bioengineering at University of Washington.